\documentclass{article}
\usepackage[a4paper, margin=1.0in]{geometry}

\usepackage{booktabs}
\usepackage{tabularx}
\usepackage{array}
\usepackage{makecell}
\usepackage{multirow}
\usepackage{graphicx}
\usepackage[table]{xcolor}
\usepackage{url}
\usepackage{tcolorbox}
\usepackage{microtype}
\usepackage{hyperref}

\definecolor{popBlue}{HTML}{2F6F9F}
\definecolor{popBlueLight}{HTML}{EAF4FB}
\definecolor{popGreenLight}{HTML}{EAF7ED}
\definecolor{popOrangeLight}{HTML}{FFF1DC}
\definecolor{popPurpleLight}{HTML}{F0ECFA}
\definecolor{popGrayLight}{HTML}{F3F4F6}
\definecolor{popBest}{HTML}{DDF5E4}

\newcolumntype{Y}{>{\raggedright\arraybackslash}X}
\newcolumntype{L}[1]{>{\raggedright\arraybackslash}p{#1}}
\newcolumntype{C}[1]{>{\centering\arraybackslash}p{#1}}
\newcommand{\primary}{\ensuremath{\bullet}}
\newcommand{\partialc}{\ensuremath{\triangle}}
\newcommand{\nonec}{\ensuremath{\circ}}
\newcommand{\metricN}{nDCG@10}
\newcommand{\metricR}{Recall@10}

\newcommand{\gain}[1]{\textcolor{popBlue}{\scriptsize #1}}

\title{Popcorn: A Configurable Benchmark for Visual Evidence in Multimodal Movie Recommendation}

\author{
    Ali Tourani\footnote{Interdisciplinary Centre for Security, Reliability, and Trust (SnT), University of Luxembourg, Luxembourg. \texttt{ali.tourani@uni.lu}},
    Fatemeh Nazary\footnote{Polytechnic University of Bari, Bari, Italy. \texttt{fatemeh.nazary@poliba.it}},
    Yashar Deldjoo\footnote{Polytechnic University of Bari, Bari, Italy. \texttt{deldjooy@acm.org}},
    Tommaso Di Noia\footnote{Polytechnic University of Bari, Bari, Italy. \texttt{tommaso.dinoia@poliba.it}},
}

\begin{document}

\maketitle

\begin{abstract}
Movies are long-form audiovisual works, yet recommender benchmarks often rely on trailers, thumbnails, or metadata. These sources differ in semantics and scalability: full movies preserve consumption-level evidence, trailers concentrate promotional highlights, and thumbnails provide sparse but catalog-scale visual signals. We present \emph{Popcorn}, a configurable benchmark for visual evidence in multimodal movie recommendation, combining title-aligned full-movie/trailer embeddings with MovieLens-linked thumbnail features encoded by modern visual and vision-language models. Popcorn standardizes modality assembly, fusion, splitting, evaluation, and LLM-augmented metadata through a single configuration contract. Experiments show that thumbnail VLMs provide strong, scalable item-side evidence, while controlled trailer/full-movie comparisons show that visual evidence sources are not interchangeable: the choice of source and fusion strategy affects ranking accuracy, coverage, diversity, and calibration. The framework is available at \href{https://github.com/RecSys-lab/Popcorn}{https://github.com/RecSys-lab/Popcorn}.
\end{abstract}

\section{Introduction and Related Resources}
\label{sec:introduction}

Movies are inherently multimodal cultural artifacts: viewers respond not only to plot and genre, but also to cast, dialogue, soundtrack, color palette, camera motion, editing rhythm, and visual style. Nevertheless, movie recommendation benchmarks often operationalize films through user--item interactions, sparse metadata, posters, or short promotional videos. Although such abstractions are practical, they obscure a fundamental modeling choice: \emph{what visual evidence is the recommender learning from?}

\begin{figure}[t]
    \centering
    \includegraphics[width=0.850\linewidth]{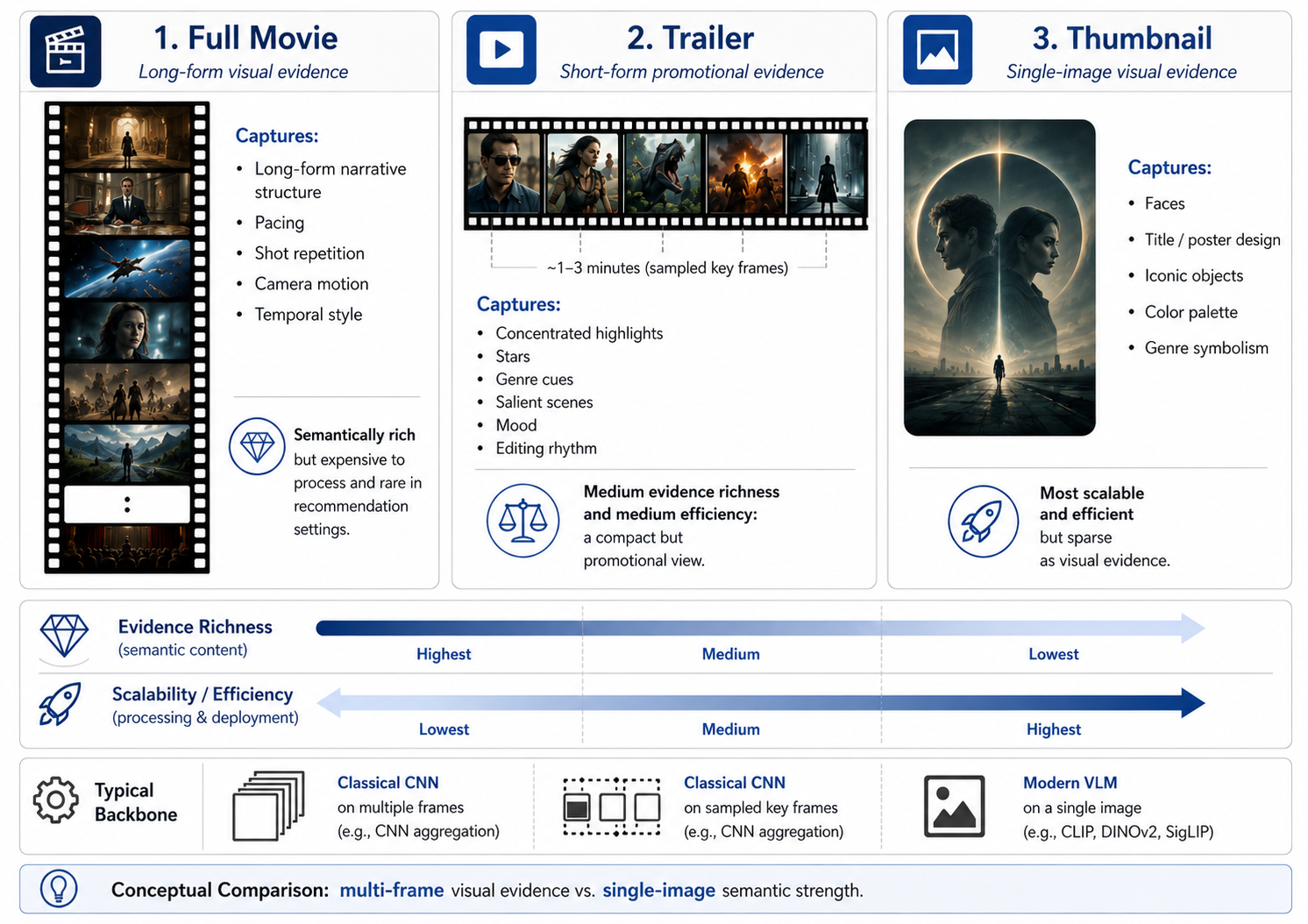}
\caption{Conceptual comparison of \textbf{full movies}, \textbf{trailers}, and \textbf{thumbnails} as visual evidence sources, contrasting \emph{multi-frame CNN evidence} from full movies/trailers with \emph{single-image VLM evidence} from thumbnails at catalog scale.}\label{fig:visual_evidence}
\end{figure}

The answer is consequential because visual evidence sources differ in semantics, availability, and computational cost (Fig.~\ref{fig:visual_evidence}). A \textbf{full movie} is closest to the consumed item and preserves narrative structure, pacing, repeated shots, camera motion, and long-form audiovisual style, but it is difficult to distribute and expensive to process. A \textbf{trailer} is compact and widely accessible, yet it is a promotional artifact that deliberately concentrates stars, genre cues, action, mood, and salient scenes. A \textbf{thumbnail} or poster is the most scalable evidence source, but it compresses a film's visual identity into a single static image, typically emphasizing faces, typography, iconic objects, color palette, and genre symbolism. This distinction also determines what visual backbones can exploit. The mainstream approach in earlier multimodal movie recommendation has been to extract CNN features from multiple video frames, typically from trailers or other sampled video data~\cite {mmtf14k,deldjoo2021recommender}. Such multi-frame CNN pipelines can capture recurring objects, textures, scene composition, lighting, and frame-level cues that act as proxies for genre, mood, or visual style. Modern vision-language models (VLMs), in contrast, can encode a single thumbnail or poster into a semantically organized image--text representation, making sparse image evidence surprisingly informative at large catalog scale. Thus, the comparison studied here is not only a backbone comparison between \emph{classical CNNs} and \emph{modern VLMs}; it is a comparison between two evidence regimes: \textbf{multiple-frame trailer/full-movie evidence with classical CNN pipelines} and \textbf{single-thumbnail semantic evidence with modern VLMs}. Popcorn aims to make this distinction explicit so that improvements can be interpreted in terms of evidence source, encoder family, scalability, and downstream recommendation behavior.

\begin{table*}[t]
    \centering
    \fontsize{5.6}{7.0}\selectfont
    \caption{Positioning Popcorn against representative multimodal recommendation resources. Symbols: \primary~= primary support, \partialc~= partial or indirect support, and \nonec~= not addressed/applicable. Popcorn is distinct in combining thumbnail, trailer, and full-movie evidence with CNN/VLM backbones, GenAI modules, Visual RAG, controlled configurations, and beyond-accuracy auditing.}
    \label{tab:related}
    \setlength{\tabcolsep}{2.4pt}
    \renewcommand{\arraystretch}{1.05}
    \begin{tabularx}{\textwidth}{L{2.65cm} C{.40cm}C{.40cm}C{.40cm}C{.40cm} C{.40cm}C{.40cm} C{.40cm}C{.40cm}C{.40cm} C{.50cm}C{.50cm} C{.55cm}C{.55cm} Y}
    \toprule
        \multirow{2}{*}{\textbf{Work / resource}} & \multicolumn{4}{c}{\textbf{Evidence}} & \multicolumn{2}{c}{\textbf{Backbone}} & \multicolumn{3}{c}{\textbf{Modalities}} & \multicolumn{2}{c}{\textbf{GenAI}} & \multicolumn{2}{c}{\textbf{Benchmark}} & \multirow{2}{*}{\textbf{Gap addressed}} \\
        \cmidrule(lr){2-5}\cmidrule(lr){6-7}\cmidrule(lr){8-10}\cmidrule(lr){11-12}\cmidrule(lr){13-14}
        & \rotatebox{90}{Thumb} & \rotatebox{90}{Trailer} & \rotatebox{90}{Full} & \rotatebox{90}{Micro} & \rotatebox{90}{CNN} & \rotatebox{90}{VLM} & \rotatebox{90}{Visual} & \rotatebox{90}{Audio} & \rotatebox{90}{Text} & \rotatebox{90}{LLM text} & \rotatebox{90}{V-RAG} & \rotatebox{90}{Config} & \rotatebox{90}{Audit} & \\
    \midrule
        Ducho / Ducho$\times$Elliot~\cite{ducho2,ducho_elliot} & \nonec & \nonec & \nonec & \nonec & \partialc & \partialc & \primary & \primary & \primary & \nonec & \nonec & \partialc & \nonec & Feature extraction toolkit; not a source-controlled movie benchmark. \\
    \midrule
        MMRec / MMSSL~\cite{mmrec,mmsl} & \nonec & \nonec & \nonec & \nonec & \partialc & \partialc & \primary & \primary & \primary & \nonec & \nonec & \partialc & \nonec & Model-centric multimodal recommendation; evidence source is not the primary variable. \\
    \midrule
        Rec-GPT4V / MMRec-LLM~\cite{recgpt,mmrec_llm} & \partialc & \nonec & \nonec & \nonec & \nonec & \primary & \primary & \nonec & \primary & \primary & \partialc & \partialc & \nonec & VLM/LLM reasoning without full-movie/trailer evidence control. \\
    \midrule
        MicroLens~\cite{microlens} & \nonec & \nonec & \nonec & \primary & \partialc & \partialc & \primary & \primary & \nonec & \nonec & \nonec & \partialc & \nonec & Large micro-video scale, but not long-form movie evidence. \\
    \midrule
        MMTF-14K~\cite{mmtf14k} & \nonec & \primary & \nonec & \nonec & \primary & \nonec & \primary & \primary & \nonec & \nonec & \nonec & \partialc & \nonec & Trailer features; no full-movie or VLM thumbnail layer. \\
    \midrule
        ViLLA-MMBench~\cite{villammbench} & \nonec & \primary & \nonec & \nonec & \partialc & \partialc & \primary & \primary & \primary & \partialc & \nonec & \primary & \partialc & Trailer-centric evaluation without full-movie alignment. \\
    \midrule
        RAG-VisualRec~\cite{ragvisualrec} & \nonec & \primary & \nonec & \nonec & \partialc & \partialc & \primary & \nonec & \primary & \primary & \primary & \primary & \partialc & Visual RAG for trailers; no controlled thumbnail/trailer/full comparison. \\
    \midrule
        \rowcolor{popBest}
        \textbf{Popcorn} & \primary & \primary & \primary & \nonec & \primary & \primary & \primary & \primary & \primary & \primary & \primary & \primary & \primary & Source-controlled visual evidence benchmark and reproducible pipeline. \\
    \bottomrule
    \end{tabularx}
\end{table*}

\begin{figure}[t]
    \centering
    \includegraphics[width=.98\linewidth]{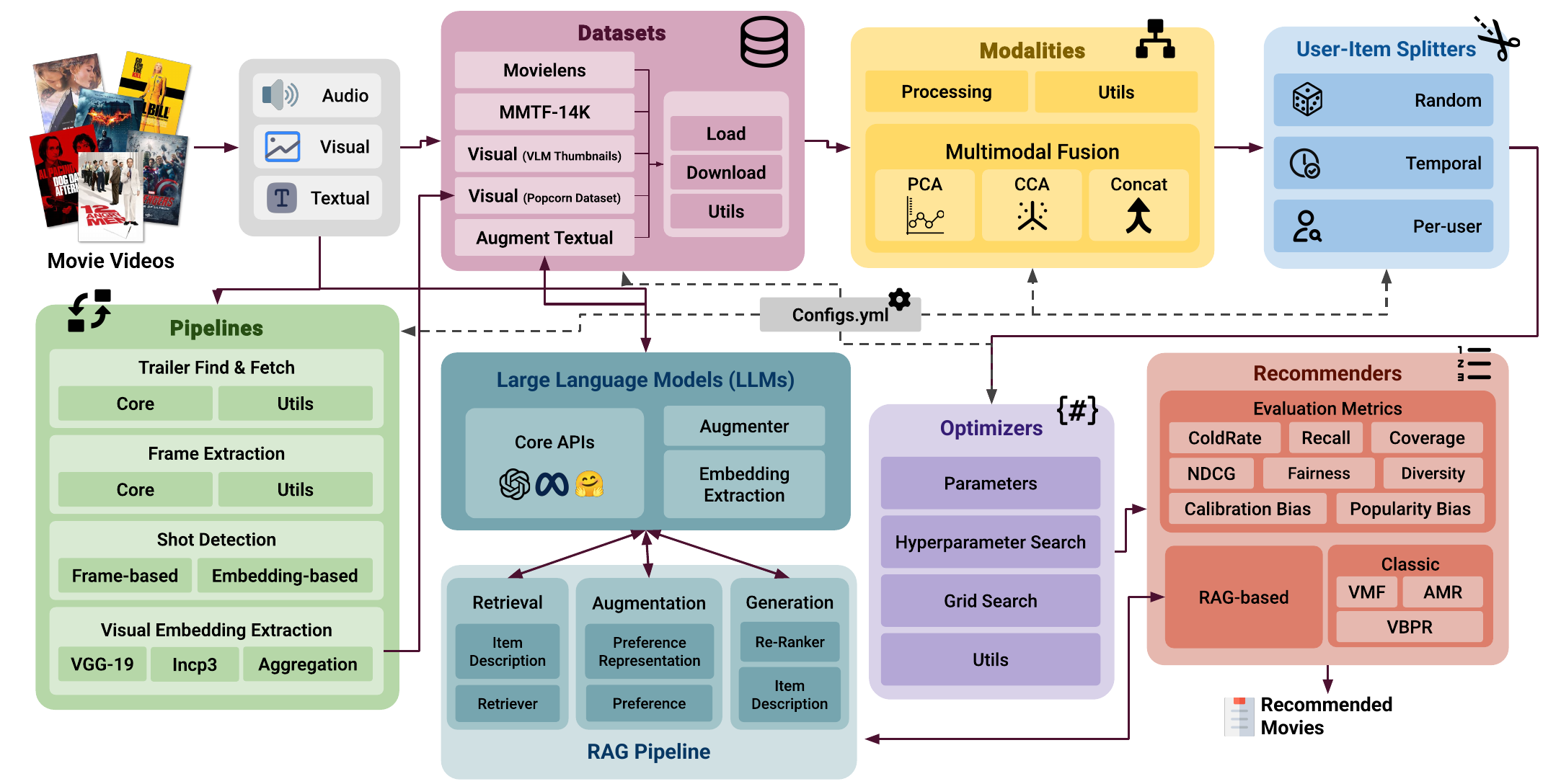}
    \caption{Popcorn architecture. A single configuration controls evidence loading, visual/audio/text pipelines, optional fusion (Concat/PCA/CCA), split construction, training, HPO, metric export, LLM-based enrichment, and Visual RAG reranking or explanations.}
    \label{fig:pipeline}
\end{figure}

\paragraph{Related resources and gap.}
As summarized in Table~\ref{tab:related}, existing multimodal recommendation resources provide important foundations but do not isolate the source of visual evidence as the central experimental variable. Feature toolkits such as Ducho and Ducho$\times$Elliot support multimodal extraction and integration~\cite{ducho2,ducho_elliot}, while MMRec and MMSSL focus on broader multimodal model benchmarking~\cite{mmrec,mmsl}. Movie and video resources such as MMTF-14K, MicroLens, ViLLA-MMBench, and RAG-VisualRec contribute trailer features, micro-video scale, or LLM/RAG-oriented protocols~\cite{mmtf14k,microlens,villammbench,ragvisualrec}. However, these resources do not jointly provide evidence for thumbnails, trailers, and full-movie data; CNN and VLM backbones; multimodal fusion; GenAI components; configuration-controlled evaluation; and beyond-accuracy auditing. This leaves open a basic question: whether conclusions drawn from trailer features transfer to full movies, and how static thumbnail evidence encoded by modern VLMs compares with multi-frame CNN evidence under a shared recommendation protocol.

We introduce \textbf{Popcorn}, a resource and configurable benchmark for controlled visual-evidence evaluation in multimodal movie recommendation. Rather than proposing a new recommender architecture, Popcorn releases complementary \emph{data resources} and a \emph{software pipeline}: (i) title-aligned full-movie/trailer evidence for 274 movies, provided as derived frame-level, shot-level, and pooled embeddings, with frame-level representations sampled at 1 FPS; (ii) a MovieLens-linked thumbnail layer covering approximately 65K titles, organized into 13 image packs and encoded with six modern visual/VLM backbones, yielding more than 300K visual embeddings; and (iii) a configuration-driven multimodal pipeline for evidence loading, fusion, training, evaluation, LLM augmentation, and Visual RAG. The benchmark allows researchers to vary evidence source, backbone, fusion, augmentation, recommender, and evaluation setting while keeping the downstream protocol fixed. Our contributions are:

\begin{itemize}
    \item \textbf{Visual-evidence benchmark (\S \ref{sec:experiments} - RQ1).}
    Popcorn frames multimodal movie recommendation as a controlled visual-evidence benchmark, directly comparing \emph{single-thumbnail VLM evidence} with \emph{multi-frame CNN evidence} from trailers and full movies under a fixed split, recommender, and evaluation protocol.

    \item \textbf{Released aligned-video and thumbnail evidence layers.}
    Popcorn releases derived embeddings for 274 title-aligned full movies and trailers at frame, shot, and pooled granularities, encoded with classical CNN backbones including Inception-v3 \cite{incp} and VGG-19 \cite{vgg19}. It further provides a scalable thumbnail/VLM layer linking approximately 65K MovieLens-25M titles to poster evidence encoded with CLIP~\cite{clip}, OpenCLIP~\cite{openclip}, DINOv2-base/large~\cite{dinov2}, SigLIP-base~\cite{siglip}, and SigLIP2-base~\cite{siglip2}.

    \item \textbf{Auditable fusion and augmentation (\S \ref{sec:experiments} - RQ2).}
    Popcorn records modality choices, PCA/CCA settings, text-augmentation state, split configuration, recommender, and exported metrics. Experiments use representative multimedia recommenders---Visual Bayesian Personalized Ranking (VBPR) \cite{vbpr}, Adversarial Multimedia Recommendation (AMR) \cite{amr}, and Visual Matrix Factorization (VMF) \cite{vmf}---to make fusion and LLM-augmentation effects explicit.

    \item \textbf{Cost-aware VLM and beyond-accuracy analysis (\S \ref{sec:experiments} - RQ3).}
    Popcorn reports recall, coverage, novelty, diversity, fairness, popularity bias, cold-rate exposure, and calibration, and relates thumbnail VLM performance to a model-size/storage proxy.
\end{itemize}

Overall, Popcorn contributes both \emph{data}---thumbnail, trailer, and full-movie evidence layers---and \emph{software}---a configurable multimodal recommendation pipeline for reproducible visual-evidence analysis.

\section{Popcorn Resource and Pipeline}
\label{sec:resource}

Popcorn is organized as a layered resource and pipeline for controlled visual-evidence benchmarking, complementing prior trailer-, micro-video-, and RAG-oriented resources~\cite{mmtf14k,microlens,villammbench,ragvisualrec}. The \emph{aligned-video layer} contains derived embeddings for 274 title-aligned full movies and official trailers, exposed at frame, shot, and pooled levels\footnote{Full-movies dataset: \url{https://huggingface.co/datasets/alitourani/Popcorn_Dataset}.}. The \emph{thumbnail/VLM layer} links approximately 65,000 MovieLens-25M titles to thumbnail or poster evidence, packaged into 13 image packs and encoded with six modern visual backbones, yielding more than 300K visual embeddings\footnote{Thumbnails: \url{https://huggingface.co/datasets/alitourani/movielens-25m-thumb}.}. The \emph{software layer} provides loaders, ID alignments, modality assembly, fusion modules, splitters, recommender wrappers, hyperparameter search, metric export, and optional LLM/RAG components. Full movies are released as derived embeddings rather than raw videos; users with lawful access can recompute features through the pipeline\footnote{Framework: \url{https://github.com/RecSys-lab/Popcorn}.}.

Let $U$ be users, $I$ movies, and $R\subseteq U\times I$ observed feedback. For item $i$, Popcorn makes visual evidence explicit as $e\in\{\mathrm{thumb},\mathrm{trailer},\mathrm{full}\}$. Given backbone $b$, granularity $g$, and pooling operator $p$, the visual representation is
\begin{equation}
\mathbf{x}_{i}^{(e,b,g,p)} = p\left(\{\psi_b(v): v\in\mathcal{V}_{i}^{(e,g)}\}\right),
\end{equation}
where $\mathcal{V}_{i}^{(e,g)}$ is a singleton image for thumbnails or a frame/shot sequence for video. Optional text and audio vectors are denoted $\mathbf{t}_i$ and $\mathbf{a}_i$. A fusion operator $\phi$ constructs $\mathbf{z}_i=\phi(\mathbf{x}_{i}^{(e,b,g,p)},\mathbf{t}_i,\mathbf{a}_i)$, where $\phi$ may be identity, concatenation, PCA, CCA, or rank aggregation.

A run is identified by $(D,I,e,b,g,p,\mathcal{M},\phi,f_\theta,s,K)$, specifying dataset, item universe, evidence source, backbone, granularity, modality set, fusion, recommender, split, and cutoff. The toolkit exports the resolved configuration, recommendation lists, and metrics, making ablations reproducible. LLM modules are optional: item enrichment expands sparse metadata into descriptions whose embeddings can be fused with visual vectors; profile enrichment summarizes interaction histories; and Visual RAG\footnote{Visual RAG is currently a planned extension of Popcorn. The Visual-RAG pipeline has been implemented and evaluated separately in RAG-VisualRec~\cite{ragvisualrec}.} injects retrieved frames, shots, or thumbnails with provenance into the LLM context for auditable reranking and explanations.

\begin{table*}[t]
    \centering
    \fontsize{5.0}{7.0}\selectfont
    \caption{Popcorn benchmark dashboard. Panel A reports the larger thumbnail/VLM slice with VBPR; $\Delta$ is relative to the MMTF-14K CNN visual baseline (0.222 \metricN, 0.203 \metricR). Panel B reports the aligned-video slice; metric pairs are trailer/full-movie, and $\Delta$ is the relative advantage of the winning source.}
    \label{tab:benchmark}
    \renewcommand{\arraystretch}{1.00}
    \begin{tabularx}{\textwidth}{C{.28cm} C{.95cm} C{.43cm} L{2.6cm} | C{1.2cm} | C{1.0cm} | C{1.0cm} | C{1.6cm} | Y}
    \toprule
    \textbf{ID} & \textbf{Source} & \textbf{Mod.} & \textbf{Encoder / features} & \textbf{Fusion} & \textbf{\makecell{nDCG\\@10}} & \textbf{\makecell{Recall\\@10}} & \textbf{$\Delta$} & \textbf{Interpretation} \\
    \midrule
    \multicolumn{9}{l}{\textbf{Panel A: larger thumbnail/VLM benchmark, $|I|\approx14$K, model = VBPR}} \\
    \midrule
    A0 & Trailer & V & MMTF-14K CNN & none & 0.222 & 0.203 & reference & Older trailer-CNN visual baseline. \\
    \rowcolor{popGrayLight}
    A1 & Audio & A & MMTF-14K BLF & none & 0.237 & 0.215 & \gain{+6.8/+5.9} & Audio side information is competitive with older CNN visual features. \\
    \rowcolor{popOrangeLight}
    A2 & Text & T & LLaMA text, text\_aug=true & none & 0.240 & 0.221 & \gain{+8.1/+8.9} & Generated textual context can be useful if logged with prompts and embeddings. \\
    \rowcolor{popBlueLight}
    A3 & Thumb & V & CLIP & none & 0.254 & 0.235 & \gain{+14.4/+15.8} & Static VLM features exceed the older CNN visual baseline. \\
    \rowcolor{popBlueLight}
    A4 & Thumb & V & DINOv2-base & none & 0.243 & 0.224 & \gain{+9.5/+10.3} & Self-supervised image features provide a strong static visual signal. \\
    \rowcolor{popBlueLight}
    A5 & Thumb & V & DINOv2-large & none & 0.248 & 0.226 & \gain{+11.7/+11.3} & Larger DINOv2 improves over base but remains below SigLIP-base. \\
    \rowcolor{popBlueLight}
    A6 & Thumb & V & OpenCLIP & none & 0.250 & 0.227 & \gain{+12.6/+11.8} & Contrastive VLM features remain robust. \\
    \rowcolor{popBlueLight}
    A7 & Thumb & V & SigLIP2-base & none & 0.250 & 0.228 & \gain{+12.6/+12.3} & Strong VLM feature, slightly below SigLIP-base here. \\
    \rowcolor{popBest}
    A8 & Thumb & V & SigLIP-base & none & \textbf{0.269} & \textbf{0.262} & \gain{+21.2/+29.1} & Best visual-only row; demonstrates the value of the thumbnail/VLM scale layer. \\
    \rowcolor{popPurpleLight}
    A9 & Fuse & V+T & SigLIP-base & PCA \tiny{(var. 0.9)} & 0.242 & 0.240 & \gain{+9.0/+18.2} & Fusion is not automatically beneficial; SigLIP visual has higher \metricN. \\
    \rowcolor{popPurpleLight}
    A10 & Fuse & V+T & SigLIP-base & CCA \tiny{(comp. 40)} & 0.268 & 0.261 & \gain{+20.7/+28.6} & CCA nearly matches the best visual-only result without surpassing it. \\
    \midrule
    \multicolumn{9}{l}{\textbf{Panel B: aligned trailer/full-movie benchmark, 274 titles; metric pairs are trailer/full-movie}} \\
    \midrule
    B1 & T/F & V & VBPR; Inception-v3 agg. max & none & 0.433/0.413 & 0.575/0.552 & Trailer \gain{+4.8/+4.2} & Trailer visual-only evidence is higher under the fixed title set. \\
    \rowcolor{popPurpleLight}
    B2 & T/F & V+T & VBPR; text+visual & CCA \tiny{(comp. 40)} & 0.444/0.436 & 0.579/0.573 & Trailer \gain{+1.8/+1.0} & Fusion narrows but does not reverse the source gap. \\
    B3 & T/F & V & AMR; Inception-v3 agg. max & none & 0.339/0.298 & 0.468/0.411 & Trailer \gain{+13.8/+13.9} & Trailer visual-only evidence is substantially higher. \\
    \rowcolor{popPurpleLight}
    B4 & T/F & V+T & AMR; text+visual & CCA \tiny{(comp. 40)} & 0.425/0.434 & 0.555/0.564 & Full \gain{+2.1/+1.6} & Fusion reverses the source ordering. \\
    B5 & T/F & V & VMF; Inception-v3 agg. max & none & 0.281/0.266 & 0.395/0.382 & Trailer \gain{+5.6/+3.4} & Trailer is slightly higher in the visual-only setting. \\
    \rowcolor{popPurpleLight}
    B6 & T/F & V+T & VMF; text+visual & CCA \tiny{(comp. 40)} & 0.275/0.285 & 0.385/0.391 & Full \gain{+3.6/+1.6} & Full movie is slightly higher after fusion. \\
    \bottomrule
    \end{tabularx}
\end{table*}

\section{Benchmark Protocol}
\label{sec:protocol}

The large-catalog experiments use MovieLens-1M \cite{movielens} interactions with a 10-core filter, top-10 recommendation, and VBPR \cite{vbpr} over approximately 14K MovieLens-linked items. They compare audio, LLM-augmented text, single-thumbnail/VLM visual features, PCA/CCA fusion, and the MMTF-14K \cite{mmtf14k} trailer-CNN visual baseline. The thumbnail backbones are CLIP \cite{clip}, OpenCLIP \cite{openclip}, DINOv2-base and -large \cite{dinov2}, SigLIP-base \cite{siglip}, and SigLIP2-base \cite{siglip2}. The aligned-video experiments use the 274-title full-movie/trailer subset with Inception-v3 \cite{incp} aggregate max-pooled CNN features and report in visual-only and text+visual CCA settings.

\begin{table}[!b]
    \centering
    \caption{Example of LLM-based data augmentation, where sparse movie metadata is expanded into a concise description while fixed fields remain unchanged.}\label{tab:augmentation}
    \scriptsize
    \setlength{\tabcolsep}{2pt}
    \renewcommand{\arraystretch}{1.01}
    \begin{tabularx}{\linewidth}{L{1.6cm} L{2.7cm} Y}
    \toprule
    \textbf{Aspect} & \textbf{Before} & \textbf{After LLM augmentation} \\
    \midrule
    Title & Nixon (1995) & unchanged \\
    Genres & Drama $|$ Biography & unchanged \\
    Description & Not provided & ``Nixon (1995) explores the troubled psyche and political career of America's 37th president, delving into his strategic brilliance and moral compromises ...'' \\
    \bottomrule
    \end{tabularx}
    \label{tbl_llm_aug}
\end{table}

\paragraph{Metrics.}
We evaluate two metric groups. \emph{Accuracy metrics} include \metricN, \metricR, precision, MAP, and hit rate, computed at top-$K$ with binary relevance. \emph{Beyond-accuracy metrics} include coverage, novelty, diversity, fairness, popularity bias, cold rate, and calibration bias. For brevity, we only name these metrics here; their formal definitions and implementation details are provided in the GitHub repository. Higher values are preferred for accuracy, coverage, novelty, diversity, fairness, and cold-rate exposure, while lower values are preferred for popularity bias and calibration bias.

\paragraph{Fusion and projection settings.}
Popcorn treats PCA and CCA as configurable hyperparameters rather than fixed preprocessing defaults. The system supports full hyperparameter search over fusion choices exposed in \texttt{config.yml}, including PCA variance thresholds, CCA component counts, and CCA regularization. In the thumbnail/VLM dashboard, PCA retains 90\% of variance and CCA uses 40 canonical components for the rows in Table~\ref{tab:benchmark}. In the aligned-video grid, the best-reported CCA configurations use 40 components with regularization parameter $\lambda=0.01$. Each exported run stores the resolved fusion method, PCA threshold, CCA dimensionality, and regularization value for reproducibility.

\paragraph{LLM data augmentation.}
Text augmentation is controlled by \texttt{text\_aug}. When enabled, an LLM expands sparse item metadata such as title, genres, tags, and missing plot descriptions into a concise paragraph describing plot, themes, style, and salient entities. The resulting text (Table~\ref{tbl_llm_aug}) is embedded by the selected backend (OpenAI, SentenceTransformer, or LLaMA-family) and can be used on its own or fused with visual/audio vectors. Augmented descriptions, prompts, embeddings, recommendation lists, and metrics can be exported for audit.

\section{Experiments and Discussion}
\label{sec:experiments}

We organize the discussion around three experimental questions: \textbf{RQ1} asks how far a single thumbnail encoded by a VLM can go compared with multi-frame CNN evidence from trailers or full movies; \textbf{RQ2} asks whether gains come from multimodal fusion or from LLM data augmentation, and how these choices affect beyond-accuracy metrics; and \textbf{RQ3} asks how thumbnail VLM performance changes with a model-size/storage proxy.

\textbf{RQ1: visual evidence.} Table~\ref{tab:benchmark} shows that a single thumbnail encoded by modern VLMs can outperform the older MMTF-14K trailer-CNN visual baseline. SigLIP-base reaches \metricN=0.269 and \metricR=0.262, corresponding to gains of 21.2\% and 29.1\%. The result should not be interpreted as thumbnails fully representing films: thumbnails cannot observe pacing, repeated shots, or narrative progression. Rather, they provide a strong catalog-scale semantic signal. On the aligned 274-title slice, trailers remain stronger than full movies in visual-only settings for VBPR, AMR, and VMF, which is plausible because trailers concentrate recommendation-salient highlights. After CCA fusion, the gap narrows for VBPR and reverses for AMR and VMF, showing that trailers and full movies are not interchangeable.

\begin{figure}[t]
    \centering
    \includegraphics[width=.50\textwidth]{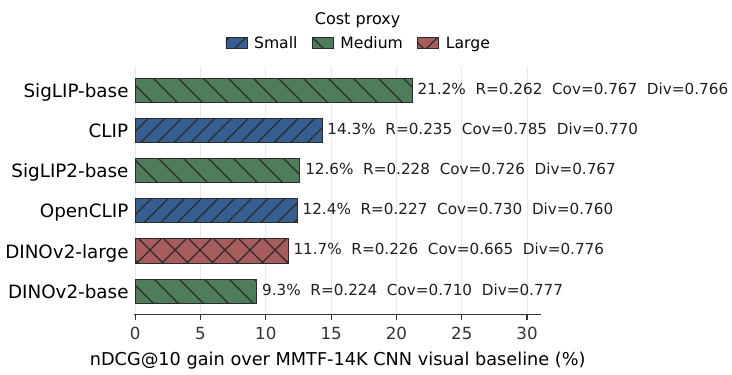}
    \caption{Thumbnail VLM trade-offs: \metricN{} gain over the MMTF-14K CNN baseline, with labels for recall, coverage, and diversity. Colors denote model-tier cost proxies.}
    \label{fig:vlmtradeoff}
\end{figure}

\textbf{RQ2: fusion versus data augmentation.} Fusion helps in some settings but is not a monotonic improvement. In Panel~A, SigLIP-base visual-only is slightly stronger than CCA on \metricN{} (0.269 vs. 0.268), while CCA increases coverage from 0.767 to 0.918 but lowers diversity from 0.766 to 0.749 and raises calibration bias from 2.901 to 3.125. PCA is weaker on accuracy (0.242 \metricN) despite retaining reasonable diversity. The aligned-video results show the same pattern: VBPR trailer CCA improves over visual-only from 0.433 to 0.444 \metricN, and AMR full-movie CCA improves over text-only from 0.378 to 0.434 \metricN and from 0.506 to 0.564 \metricR. Beyond-accuracy values show the cost of this gain: for AMR full movies with OpenAI text and no augmentation, CCA reaches 0.434 \metricN{} and 0.564 \metricR{}, but diversity drops to 0.742, below visual-only 0.773 and text-only 0.763. Data augmentation is also model-dependent: VBPR benefits modestly from LLaMA augmentation in the fused full-movie row (0.436 vs. 0.431 \metricN{} without augmentation), whereas AMR's strongest CCA row uses OpenAI text without augmentation. Full grids are provided in the GitHub repository.\footnote{Due to space limitations, for this RQ we report a compact subset of the original experimental results; the full grids and beyond-accuracy values are provided in the project repository: \url{https://recsys-lab.github.io/Popcorn/}.}

\textbf{RQ3: VLM cost versus performance.} Figure~\ref{fig:vlmtradeoff} modernizes the attached bar plot by removing calibration and storage text while retaining recall, coverage, and diversity. SigLIP-base has the best accuracy and recall, but CLIP has the highest coverage (0.785), and DINOv2-base/large has the highest diversity (0.777/0.776). The color-coded tiers show that performance is not monotonic with the cost proxy: the medium SigLIP-base is strongest in accuracy, the small CLIP is strongest in coverage, and the large DINOv2-large is not the best overall. Backbone selection should therefore depend on the intended deployment objective rather than model size alone.

\section{Conclusion and Limitations}
\label{sec:conclusion}
We presented Popcorn, a configurable benchmark for visual evidence in multimodal movie recommendation. Popcorn separates thumbnails, trailers, and full movies while logging the backbone, fusion, augmentation, split, recommender, and evaluation settings needed for reproducible ablations. Results show that modern VLM thumbnails improve over older CNN visual baselines, while trailer/full-movie evidence, fusion, and augmentation affect both accuracy and beyond-accuracy behavior.

The main limitations are scale, access, and offline evaluation. Full movies are released as derived embeddings; the aligned full-movie subset is smaller than the thumbnail layer, and LLM augmentation remains sensitive to model and prompt choices. Future work will extend Popcorn with larger lawful-access full-movie collections, stronger temporal encoders, audio-centric ablations, Visual RAG integration, user studies, and online evaluation.

\bibliographystyle{abbrv}
\bibliography{refs}

\end{document}